\documentclass[aps, twocolumn, prb,showpacs,amsmath,amssymb,superscriptaddress,letterpaper]{revtex4}
\usepackage{times}
\usepackage{amsfonts}
\usepackage{mathrsfs}
\usepackage{graphicx}
\usepackage{dcolumn}
\usepackage{bm}
\usepackage{color}

\usepackage[colorlinks,bookmarks=false,citecolor=blue,linkcolor=red,urlcolor=blue]{hyperref}
\def\be{\begin{equation}}       \def\ee{\end{equation}}
\def\bea{\begin{eqnarray}}      \def\eea{\end{eqnarray}}

\begin{document}
\title{An ab-initio study of circular photogalvanic effect in  chiral multifold semimetals}

\author{Congcong Le}
\affiliation{Max Planck Institute for Chemical Physics of Solids, 01187 Dresden, Germany}

\author{Yang Zhang}
\affiliation{Department of Physics, Massachusetts Institute of Technology, Cambridge, Massachusetts 02139, USA}

\author{Claudia Felser}
\affiliation{Max Planck Institute for Chemical Physics of Solids, 01187 Dresden, Germany}
\affiliation{Oxford Street, LISE 308 Cambridge, Massachusetts 02138, USA}

\author{Yan Sun}\email{Corresponding: ysun@cpfs.mpg.de}
\affiliation{Max Planck Institute for Chemical Physics of Solids, 01187 Dresden, Germany}

\date{\today}

\begin{abstract}
So far, the circular photogalvanic effect (CPGE) is
the only possible quantized signal in Weyl semimetals.
With inversion and mirror symmetries broken, Weyl and
multifold fermions in band
structures with opposite chiralities can
stay at different energies and generate a net
topological charge. Such kind of net topological charge
can present as a quantized signal in the circular
polarized light induced injection current.
According to current theoretical understanding,
RhSi and its counterparts are believed to be the most
promising candidate for the experimental observation
of the quantized CPGE. However, the real quantized
signal was not experimentally observed to date.
Since all the previous theoretical studies for
the quantized CPGE were based on effective model
but not realistic band structures, it
should lose some crucial details that influence
the quantized signal. The current status motives
us to perform a realistic ab-initio study for
the CPGE. Our result shows that the
quantized value is very easy to be interfered by
trivial bands related optic transitions, and
an fine tuning of the chemical potential by doping is
essential for the observation of quantized CPGE. This work
performs the first ab-initio analysis for the
quantized CPGE based on realistic electronic
band structure and provides an effective way
to solve the current problem for given materials.

\end{abstract}


\maketitle

\section{Introduction}
The band crossings with nontrivial topological invariant,
including Weyl, Dirac and unconventional fermions, have
been attracting enormous attention in condensed
matter\cite{Chiu2016,Armitage2018,Bradlyn2016}.
For Weyl and Dirac fermions\cite{HgCrSe,XG Wang,intermediate phase,multilayer weyl,xu2015,Lv2015,Weng2015,Shekhar2015,Yang2015,Xu2015,Dirac Kane,Na3Bi,Cr3As2,le2017,le2018},
they were realized experimentally in the topological
semimetal materials and further classified into the
type-I and type-II classes\cite{type2}. Unconventional
fermions, which contain three-, four-, six- and
eightfold degenerate points, are exhaustively
classified by space group symmetries in solid-state
systems with spin-orbit coupling and time-reversal
symmetry. These multifold degenerate points with
nontrivial topological numbers lead to a series of
exotic effects such as surface Fermi arcs
\cite{HgCrSe,XG Wang}, the chiral anomaly\cite{Burkov,CAE},
large anomalous Hall and spin Hall effect\cite{Enk_2017,Wangqi2017,sun2016}
and circular photogalvanic effects\cite{Zhong2016,Ma2015,Chan2017,Ma2017,Juan2017,Felix2018}.

Recently, it has been pointed out \cite{Juan2017,Felix2018}
that a single multifold degenerate point with nontrivial
invariant can lead to quantized CPGE trace, which can
directly measure the topological charge of degenerate
points. So far it is the only quantized signal exists
in topological semimetals. Those topological degenerate
points in solid-state systems always obey the
Nielsen-Ninomiya Theorem as the ground
rule\cite{Nielsen1,Nielsen2,Nielsen3}. A topological
degenerate point inevitably accompanies another
topological point with opposite charge, because the
total charges in the entire BZ must be neutralized.
If all degenerate points with opposite charges are the
same energy due to crystal symmetries, the total CPGE
trace contributed to all degenerate points is zero.
In the time-reversal and crystal symmetries, only
inversion and mirror symmetries can change the charge
sign of degenerate points, indicating that the degenerate
points with opposite charges have the same energy for the
system with inversion and mirror symmetries. Therefore,
the nonmagnetic chiral topological semimetal materials
with chiral space group, only including time-reversal
and rotational symmetries, have different energy for
degenerate points with opposite charge and can come
true quantized CPGE trace due to Pauli blocking.

In the chiral topological semimetal materials, RhSi
is the most promising candidate to exhibit quantized
CPGE trace. There are two reasons: the multifold
degenerate points and exotic Fermi arcs were confirmed
by ARPES measurements\cite{Daniel2019}; A big energy
difference between fourfold and sixfold points exist,
which is beneficial to acquire quantized CPGE trace
due to Pauli blocking\cite{Felix2018,Chang2017,Tang2017}.
Hence, RhSi gets extensive attentions in experiments.
However, theoretically predicted the quantized CPGE
\cite{Felix2018,Chang2017} is not observed in
experiments\cite{Dylan2019}.
We notice that all the previous theoretical studies for
quantized CPGE are based on few bands effective $k\cdot p$
and tight binding models with only considering the bands
near degenerated points. Since the second order
optical response is very sensitive to the details of
bands, such kind of approximation is easy to loss some
crucial information that influence the quantized signal.
The current situation motive us to perform an ab-initio
analysis for the CPGE in RhSi based on realistic
electronic band structure by first principle calculations.


The paper is organized as follows. In Section.~\ref{S1},
the symmetry analysis of CPGE tensor in chiral topological
semimetal RhSi is presented. Then, in Section.~\ref{S2},
the difference of CPGE trace between effective model and
real materials are summarized. In Section.~\ref{S3},
we discuss CPGE trace in RhSi and other related compounds(CoSi, PdGa, PtGa, PtAl, RhSn). Finally, in Section.~\ref{S4},
we give a summary of our paper.

\section{Symmetry analysis of CPGE tensor}\label{S1}

The CPGE injection current and  CPGE tensor $\beta$
can be written as\cite{Juan2017,Sipe2000}

\begin{eqnarray}
\frac{\mathrm{d} j_{i}}{\mathrm{d} t}=\beta_{i j}(\omega)\left[\vec{E}(\omega) \times \vec{E}^{*}(\omega)\right]_{j},\nonumber
\\
\beta_{i j}(\omega)=\frac{\pi e^{3}}{\hbar^2 V}  \sum_{\vec{k}, n, m} f_{n m}^{\vec{k}} \Delta_{\vec{k}, n m}^{i} R^{j}_{\vec{k},nm} \delta(\hbar \omega-E_{\vec{k}, m n})\label{beta}
\end{eqnarray}

Where $\vec{E}^{*}(\omega)=\vec{E}(-\omega)$ is the
electric field of circularly polarized light, and $i$
and $j$ index are the direction of current and circular
polarized light respectively.
$R^{j}_{\vec{k},nm}=\epsilon_{j k l}r_{\vec{k}, n m}^{k} r_{\vec{k}, m n}^{l}$, $E_{\vec{k}, m n}=E_{\vec{k}, m}-E_{\vec{k}, n}$
and $f_{n m}^{\vec{k}}=f_{n}^{\vec{k}}-f_{m}^{\vec{k}}$
are difference between band energies and Fermi-Dirac
distributions, $\Delta_{\vec{k}, n m}^{i}=\partial_{k_{i}} E(\vec{k})_{n m}$, and $r^{i}_{\vec{k}, n m}=i\left\langle m(\vec{k})\left|\partial_{k_{i}}\right| n(\vec{k})\right\rangle$.

The relation of $\Delta_{\vec{k}, n m}^{i}$, $R^{i}_{\vec{k},nm}$
and $r^{i}_{\vec{k}, n m}$ between $\vec{k}$ and $g\vec{k}$ is given by

\begin{eqnarray}
\Delta_{g\vec{k}, n m}^{i}&=&\frac{\partial E_{n m}(g\vec{k})}{\partial k_{i}}=\sum_{i^{\prime}}\frac{\partial (g\vec{k})_{i^{\prime}}}{\partial k_{i}}\Delta_{\vec{k}, n m}^{i^{\prime}},\nonumber
\\
r^i_{g\vec{k}, n m}&=&i\langle m(g\vec{k})|\frac{\partial}{\partial {k_i}}| n(g\vec{k})\rangle=\sum_{i^{\prime}}\frac{\partial (g\vec{k})_{i^{\prime}}}{\partial k_{i}}r^{i^{\prime}}_{\vec{k}, n m},\nonumber
\\
R^{i}_{g\vec{k},nm}&=&\epsilon_{i k l}[\sum_{k^{\prime}}\frac{\partial (g\vec{k})_{k^{\prime}}}{\partial k_{k}}r^{k^{\prime}}_{\vec{k}, n m}] [\sum_{l^{\prime}}\frac{\partial (g\vec{k})_{l^{\prime}}}{\partial k_{l}}r^{l^{\prime}}_{\vec{k}, n m}],\nonumber
\\
&=&\epsilon_{i k l}\sum_{k^{\prime}, l^{\prime}}\frac{\partial (g\vec{k})_{k^{\prime}}}{\partial k_{k}}\frac{\partial (g\vec{k})_{l^{\prime}}}{\partial k_{l}}r^{k^{\prime}}_{\vec{k}, n m}r^{l^{\prime}}_{\vec{k}, n m},\nonumber
\\
&=&\epsilon_{i k l}\sum_{k^{\prime}, l^{\prime}}\frac{\partial (g\vec{k})_{k^{\prime}}}{\partial k_{k}}\frac{\partial (g\vec{k})_{l^{\prime}}}{\partial k_{l}}\epsilon_{i^{\prime} k^{\prime} l^{\prime}}\epsilon_{i^{\prime} k^{\prime} l^{\prime}}r^{k^{\prime}}_{\vec{k}, n m}r^{l^{\prime}}_{\vec{k}, n m},\nonumber
\\
&=&\epsilon_{i k l}\sum_{k^{\prime}, l^{\prime}}\frac{\partial (g\vec{k})_{k^{\prime}}}{\partial k_{k}}\frac{\partial (g\vec{k})_{l^{\prime}}}{\partial k_{l}}\epsilon_{i^{\prime} k^{\prime} l^{\prime}}R^{i^{\prime}}_{\vec{k},nm}
\end{eqnarray}

 Then, we consider the relation of $\beta_{i j}(\vec{k},\omega)$ and $\beta_{i j}(g\vec{k},\omega)$ connected by crystal symmetry $g$, which is given by

\begin{eqnarray}
&&\beta_{i j}(g\vec{k},\omega),\nonumber
\\
&=&\frac{\pi e^{3}}{\hbar V}  \sum_{k, n, m} f_{n m}^{g\vec{k}} \Delta_{g\vec{k}, n m}^{i} R^{j}_{g\vec{k},nm} \delta(\hbar \omega-E_{g\vec{k}, m n}),\nonumber
\\
&=&\frac{\pi e^{3}}{\hbar V} \epsilon_{i k l} \sum_{\vec{k}, n, m} f_{n m}^{\vec{k}} [\sum_{i^{\prime}}\frac{\partial (g\vec{k})_{i^{\prime}}}{\partial k_{i}}\Delta_{\vec{k}, n m}^{i^{\prime}}]\nonumber
\\
&&[\sum_{k^{\prime}, l^{\prime}}\frac{\partial (g\vec{k})_{k^{\prime}}}{\partial k_{k}}\frac{\partial (g\vec{k})_{l^{\prime}}}{\partial k_{l}}\epsilon_{j^{\prime} k^{\prime} l^{\prime}}R^{j^{\prime}}_{\vec{k},nm}] \delta(\hbar \omega-E_{\vec{k}, m n})\nonumber
\\
&=&\sum_{i^{\prime}, k^{\prime}, l^{\prime}}\frac{\partial (g\vec{k})_{i^{\prime}}}{\partial k_{i}}\frac{\partial (g\vec{k})_{k^{\prime}}}{\partial k_{k}}\frac{\partial (g\vec{k})_{l^{\prime}}}{\partial k_{l}}\epsilon_{j^{\prime} k^{\prime} l^{\prime}}\epsilon_{i k l}\beta_{i^{\prime} j^{\prime}}(\vec{k},\omega)
\end{eqnarray}

\begin{figure}
\centerline{\includegraphics[width=0.45\textwidth]{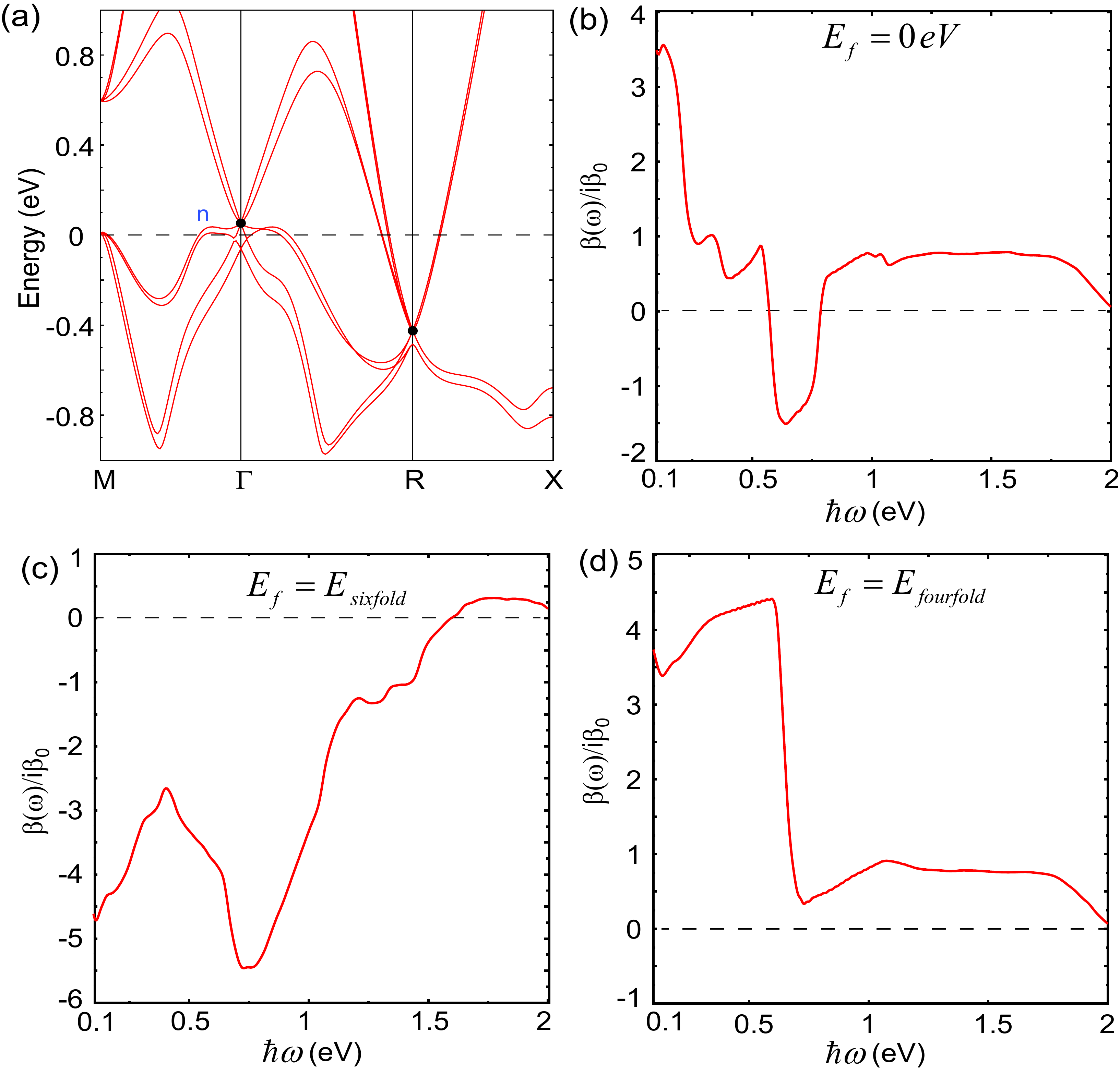}}
\caption{(color online) (a) band structure of RhSi with SOC in the paramagnetic state; black points denote fourfold and sixfold degenerate points at $\Gamma$ and M points, and the numbers n with blue color represent a band number of four degenerate points. (b)-(d) show trace of the CPGE tensor as  a function of frequency $\hbar\omega$ by DFT calculations with corresponding Fermi level $E_{f}=0~eV, E_{sixfold}, E_{fourfold}$, respectively.
\label{band} }
\end{figure}

Hence, the nonsymmorphic symmetries, which are screw
axis only in RhSi, have the same as symmorphic
symmetries for the CPGE tensor. It is clear
that the CPGE tensor $\beta_{i j}(\omega)$ of
system with inversion symmetry is zero due to
$\beta_{i j}(\vec{k},\omega)=-\beta_{i j}(-\vec{k},\omega)$,
and the materials with non-zero CPGE tensor should belong
to the gyrotropic point groups without inversion.

The crystal structure of RhSi has nonsymmorphic
space group G=P2$_1$3 (No.198), and  the quotient
group G/T is specified by 12 symmetry operations,
where we denote T as the translation group with
respect to the unit cell. In 16 symmetry operations,
nonsymmorphic symmetry operations are
$\tilde{C}_{2x}=\{C_{2x}|1/2, 1/2, 0\}$, $\tilde{C}_{2y}=\{C_{2y}|0, 1/2, 1/2\}$,
and $\tilde{C}_{2z}=\{C_{2y}|1/2, 0, 1/2\}$.
Under crystal symemtries $g=C_{2x/y/z}$, the
relation of $\beta_{i j}(\vec{k},\omega)$ and
$\beta_{i j}(g\vec{k},\omega)$ can be written as

\begin{eqnarray}
\beta_{i i}(g\vec{k},\omega)&=&\beta_{i i}(\vec{k},\omega)~~(i=j)\nonumber
\\
\beta_{i j}(g\vec{k},\omega)&=&-\beta_{i j}(\vec{k},\omega)~~(i\neq j)
\end{eqnarray}

Due to $C_{2x/y/z}$ rotational symemtries,
$\beta_{i j}(\omega)$ ($i\neq j$) is zero
and $\beta_{i i}(\omega)$ is non-zero. We will
focus on diagonal CPGE tensor $\beta_{i i}(\omega)$.
Under crystal symemtries $g=C_{3,111}$, the relation
of $\beta_{i j}(\vec{k},\omega)$ and
$\beta_{i j}(g\vec{k},\omega)$ can be given by

\begin{eqnarray}
\beta_{xx}(\vec{k},\omega)&=&\beta_{zz}(C_{3,111}\vec{k},\omega),~~\beta_{yy}(\vec{k},\omega)=\beta_{xx}(C_{3,111}\vec{k},\omega)
,\nonumber
\\
\beta_{zz}(\vec{k},\omega)&=&\beta_{yy}(C_{3,111}\vec{k},\omega)
\end{eqnarray}

\section{Difference between effective model and real materials}\label{S2}

From the effective models for multifold degenerate
points with linear dispersion, the exact quantization
of the CPGE trace is predicted in a certain frequency
range\cite{Juan2017,Felix2018}. The following is a brief
summary of derivation about quantized CPGE
trace \cite{Juan2017,Felix2018}. The CPGE
trace can be rewritten as

\begin{eqnarray}
\beta(\omega)=\frac{4\pi^2 \beta_0}{ V}  \sum_{\vec{k}, i, n, m} f_{n m}^{\vec{k}} \Delta_{\vec{k}, n m}^{i} R^{i}_{\vec{k},nm} \delta(\hbar \omega-E_{\vec{k}, m n})
\end{eqnarray}

Where $\beta_0=\frac{\pi e^{3}}{h^2}$, and the relation
between $\vec{R}_{\vec{k},n m}$ and Berry curvature
is $\vec{\Omega}_{\vec{k},n}=i \sum_{m \neq n} \vec{R}_{\vec{k},n m}$.

The topological property of degenerate points can be characterized by
the Chern number

\begin{eqnarray}
\mathcal{C}_{n}=\frac{1}{2 \pi} \oint_{S} \vec{\Omega}_{n}(\vec{k}) \cdot d \vec{S}_n
\end{eqnarray}

Where $\vec{S}_n$ is a closed surface of band $n$ enclosing
the degenerate points, and $\vec{\Omega}_{n}(\vec{k})=\boldsymbol{\nabla}_{\vec{k}} \times\langle\psi_{n}(\vec{k})|i \boldsymbol{\nabla}_{\vec{k}}| \psi_{n}(\vec{k})\rangle$
is the Berry curvature of band $n$.

In spherical coordinates, The CPGE trace becomes

\begin{eqnarray}
&&\beta(\omega)\nonumber
\\
&=&4 \pi^{2} \beta_{0} \sum_{i,k, n, m} \int \frac{k^{2} \mathrm{d} k d \Omega}{(2 \pi)^{3}}  \partial_{\vec{k}} E_{\vec{k},n m}\cdot\vec{R}_{\vec{k},n m} \delta(\hbar \omega-E_{\vec{k}, m n})\nonumber
\\
&=&4 \pi^{2} \beta_{0} \sum_{n, m}\int \frac{k^{2} \mathrm{d} k d \Omega}{(2 \pi)^{3}}  \partial_{k} E^{\hat{k}}_{k,n m}  R^{\hat{k}}_{k,n m} \delta(\hbar \omega-E_{\vec{k}, m n})\nonumber
\\
&=&4 \pi^{2} \beta_{0} \sum_{n, m}\int \frac{ \mathrm{d} (E_{k,n m}) d \Omega}{(2 \pi)^{3}}   k^{2} R^{\hat{k}}_{k,n m} \delta(\hbar \omega-E_{\vec{k}, m n})\nonumber
\\
&=&4 \pi^{2} \beta_{0} \sum_{n, m}\int \frac{  d \Omega}{(2 \pi)^{3}}   k^{2}(\hbar \omega) R^{\hat{k}}_{n m}(\hbar \omega) \nonumber
\\
&=&4 \pi^{2} \beta_{0} \sum_{n, m}\int d S^{\hat{k}}_{n m}  R^{\hat{k}}_{n m},\nonumber
\\
&=&4\pi^2 \beta_0 \sum_{n, m} \int d \vec{S}_{n m} \cdot \vec{R}_{n m}
\end{eqnarray}

Where $\partial_{\vec{k}} E_{\vec{k},n m}=\partial_{k} E^{\hat{k}}_{k,n m} \hat{k}+\frac{1}{k} \partial_{\theta} E^{\hat{\theta}}_{k,n m} \hat{\theta}+\frac{1}{k \sin \theta} \partial_{\phi} E^{\hat{\phi}}_{k,n m} \hat{\phi}$ and $\vec{R}_{\vec{k},n m}=R_{k,n m}^{\hat{k}}\hat{k}+R_{k,n m}^{\hat{\theta}}\hat{\theta}+R_{k,n m}^{\hat{\phi}}\hat{\phi}$, $\vec{R}_{nm}$
has only the radial component in spherical coordinates
for multifold degenerate points with linear dispersion.
For a given frequency $\omega$, the delta function and
Fermi-Dirac distributions select a surface $\vec{S}_{nm}$
in the $\vec{k}$ space where $E_{\vec{k}, n m}=\hbar\omega$,
and $d\vec{S}$ denotes the oriented surface element
normal to $\vec{S}$. Hence the CPGE trace is physically
understood as the Berry flux penetrating through
$\vec{S}$. For a single type-I Weyl point, when the
Fermi level is located at the Weyl point, the CPGE
trace is $\beta(\omega) =i C \beta_{0}$ where $C$ is
Chern number of the occupied band. Similar to
type-I Weyl points, other multifold fermions can
also obtain quantized CPGE trace according to the
above derivation.

The above results are based on the effective $k\cdot p$
models of multifold fermions. However, for real nonmagnetic
chiral topological semimetal materials,
there are some factors that affect quantized CPGE trace:

(1) Except for the bands that form the degenerate
points, there are many extra bands in real materials,
which can induce corrections of relation between
$\vec{R}$ and $\vec{\Omega}_{n}$ compared with
that of the effective model.

(2) The condition for quantization is that the
band structures near multifold fermion have
linear dispersion, indicating that higher-order
band dispersions can modify the CPGE trace.

(3) The nontrivial band structures that form
multifold fermions can contribute quantized
CPGE trace, while other trivial band structures
are allowed optical transitions and can provide CPGE.

\section{Ab initio analysis of circular photogalvanic effect(CPGE)}\label{S3}

\begin{figure*}
\centerline{\includegraphics[width=0.8\textwidth]{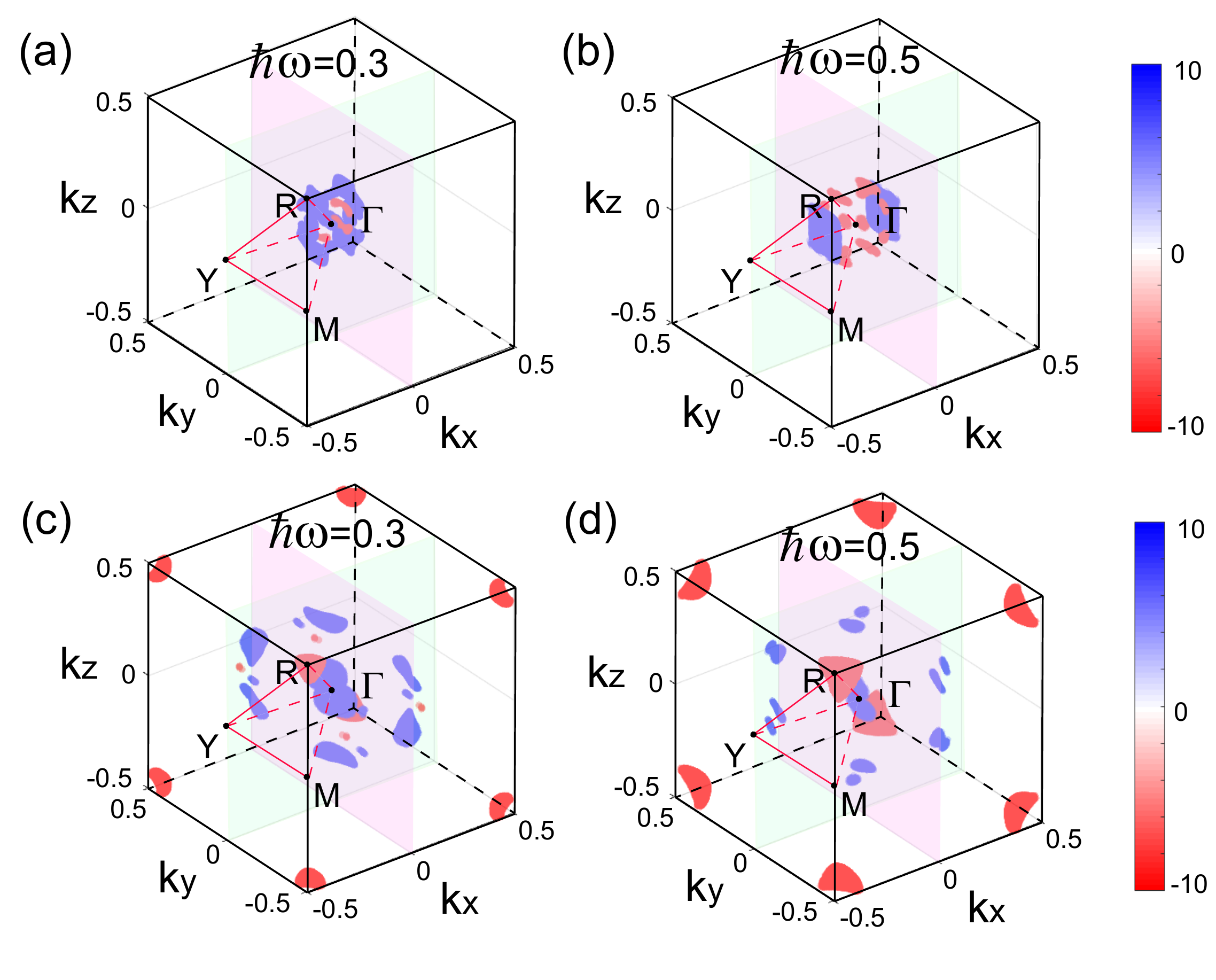}}
\caption{(color online) (a)-(b) Momentum distribution of CPGE tensor $\beta_{xx}$ from all bands without adjusting Fermi level at frequecy $\hbar\omega=0.3$ and $0.5~ eV$. (c)-(d) Momentum distribution of CPGE tensor $\beta_{xx}$ in the Brillouin zone from all bands with Fermi level $E_f=E_{sixfold}$ at frequecy $\hbar\omega=0.3$ and $0.5 eV$.
\label{fig5} }
\end{figure*}

To calculate CPGE tensor, we project the $ab-initio$ DFT Bloch
wave function into high symmetric atomic-orbital-like
Wannier functions~\cite{Yates2007} with diagonal position operator, as performed in the code of the full-potential local-orbital
minimum-basis (FPLO)~\cite{Koepernik1999, perdew1996}.
For obtaining precise Wannier functions, we include the most
outside d-, s-, and p-orbital for Rh, most outside s- and p-orbital
for element Si, which guarantees the full bands overlap from
ab-initio and Wannier functions in the energy window from -10 to 10 eV.
With highly symmetric Wannier functions, we
construct an effective tight-binding model Hamiltonian and
calculate the photoconductivity by the Eqs.~\ref{beta}.

Fig.\ref{band}(a) shows the band structure of
RhSi with spin-orbital coupling (SOC) in the paramagnetic state. Near
the Fermi level, the valence and conduction
bands are mainly attributed to the Rh-4d
orbitals. A fourfold degeneracy at $\Gamma$ points
with the energy $E_{fourfold}$=50 meV is protected by point group
T and time-reversal symmetry, which can be understood
by the character tables of point group T. A sixfold
degeneracy at $R$ points with the energy $E_{sixfold}$=-400 meV
is protected by non-symmorphic symmetries
($\{C^{-1}_{3,111}|010\}$,
$\{C_{2x}|\frac{1}{2}\frac{3}{2}0\}$
and $\{C_{2y}|0\frac{3}{2}\frac{31}{2}\}$)
and time-reversal symmetry\cite{Bradlyn2016}.
Because degeneracy points at $\Gamma$ and $R$
are not symmetrically related, the energies of
these points can be different. The fourfold
degeneracy at $\Gamma$ point is described a
spin-3/2 fermion with Chern number C=3, 1, -1, -3
for four bands\cite{Chang2017,Tang2017}, and
the sixfold degeneracy at $R$ point is double
spin-1 fermion with C=2, 2, 0, 0, -2,-2 for
six bands, which satisfies Nielsen-Ninomiya
theorem\cite{Nielsen1,Nielsen2,Nielsen3}.

We firstly calculated the trace of CPGE tensor with Fermi
level lying at the charge neutral points,
which should close to the case in experimental
measurements\cite{Dylan2019}. As presented in Fig.\ref{band}(b), one
can see any quantized value in the long rang
of frequency from ~0.1 to 2.0 eV, and the plateau close to 4$\beta_0$ is absent, which consists with experimental results\cite{Dylan2019}. For explaining that the quantization disappears in Fig.\ref{band}(b), we take two frequencies as examples to calculate the local momentum distribution of CPGE tensor $\beta_{xx}$. When two degenerate points with opposite charge can simultaneously provide CPGE, the quantization will disappear and the total of CPGE is close to zero. Hence, two frequencies $\hbar\omega=0.3$ and $0.5~eV$ are chosen, where only the fourfold degenerate point can contribute CPGE. From Fig.\ref{fig5}(a)-(b), one can easily see that the trivial bands are mainly attributed to CPGE trace and the closed surface $\vec{S}$ wrapping degenerate point also absent. As discussed in Section.~\ref{S2}, the single multifold degenerate point with nontrivial topological charge can contribute to quantized CPGE trace in a certain frequency range, and the CPGE trace can be understood as the Berry curvature penetrating through closed surface $\vec{S}$ wrapping degenerate point. The results in Fig.\ref{fig5}(a)-(b) does not satisfy the conditions for obtaining quantized CPGE trace, and hence the quantization of CPGE trace with Fermi level lying at the charge neutral point can not be observed in our calculations.

To further explore the quantized CPGE trace in RhSi and provide clues for experimental research, we can dope holes and electrons in our calculations by reducing and raising the Fermi level. In doping holes the quantized CPGE trace is always absent, while the quantized platform close to 4$\beta_0$ can be acquired in a certain range in doping electrons. In the following, we will describe both cases in detail.

In doping holes in RhSi by reducing Fermi level, and the quantized platform is similar to that of Fermi
level lying at the charge neutral points and always absent in a certain range of doping holes. We will analyze the reasons by focusing on the case of Fermi level at sixfold degeneracy. Fig.\ref{band}(c) show the CPGE trace with Fermi level
$E_{f}=E_{sixfold}$.
Although the optical transitions near fourfold fermion are
forbidden due to Pauli blocking in the
frequency range from 0.1 to 0.5 eV, the
plateau close to 4$\beta_0$ contributed to
sixfold degeneracy is absent. Similar to the no doping case, the CPGE trace provided
by trivial bands can compete with that
contributed by the sixfold degeneracy
point. Fig.\ref{fig5}(c)-(d) show
momentum distribution of CPGE tensor
$\beta_{xx}$ in the Brillouin zone from
all bands with Fermi level $E_{f}=E_{sixfold}$
at frequency $\hbar\omega=0.3$ and $0.5 eV$, and
it is clear that the quantized CPGE trace is
greatly affected by trivial bands, indicating that the quantization of CPGE trace  disappears in doping holes.


\begin{figure*}
\centerline{\includegraphics[width=0.8\textwidth]{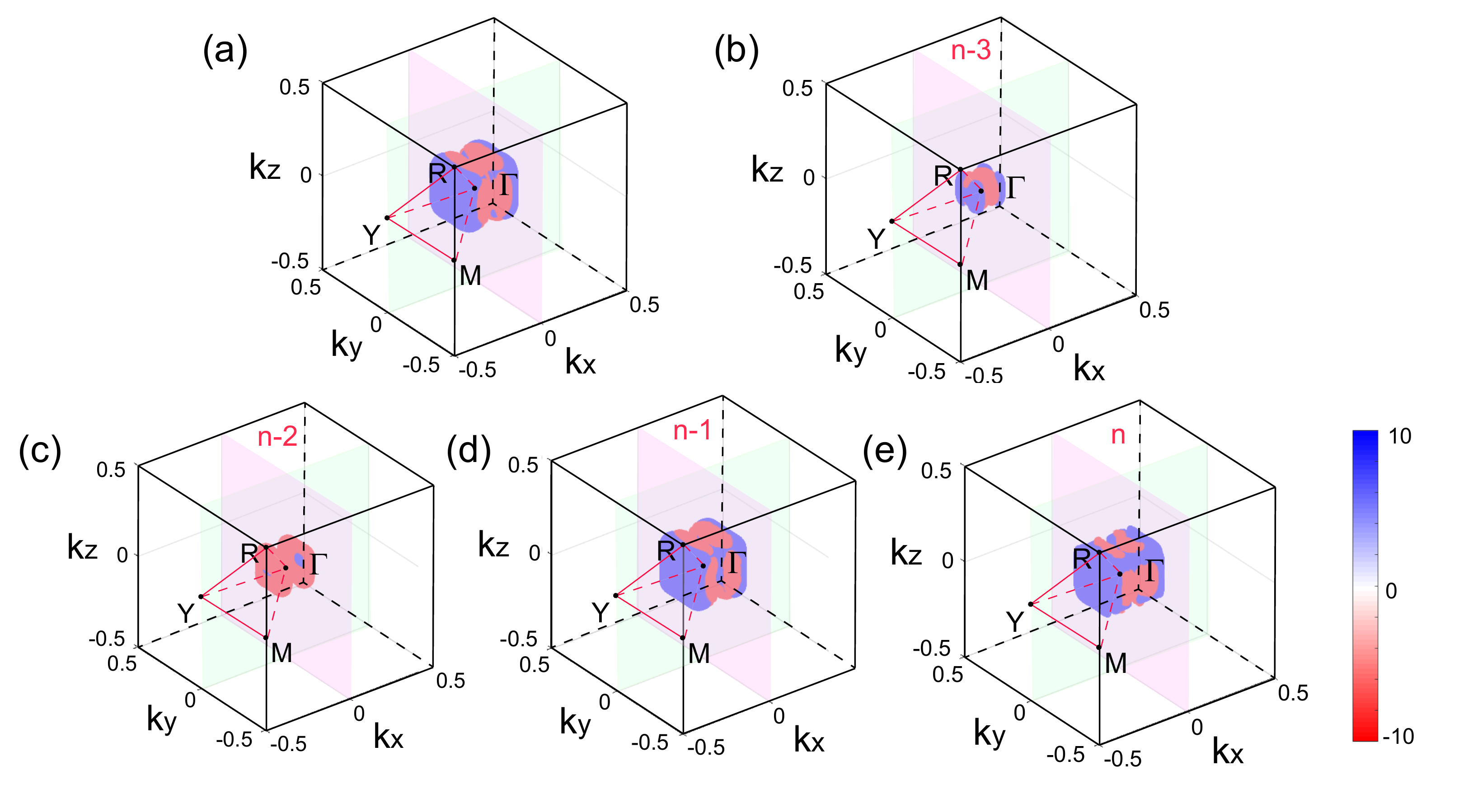}}
\caption{(color online) Momentum distribution of CPGE tensor $\beta_{xx}$ in the Brillouin zone at frequency $\hbar\omega=0.5 eV$ with corresponding Fermi level $E_{f}=E_{fourfold}$. (a) Momentum distribution from all bands with optically allowed transitions; (b)-(e) Momentum distribution from band-(n-3) to band-n, respectively.
\label{fig3} }
\end{figure*}

\begin{figure*}
\centerline{\includegraphics[width=0.8\textwidth]{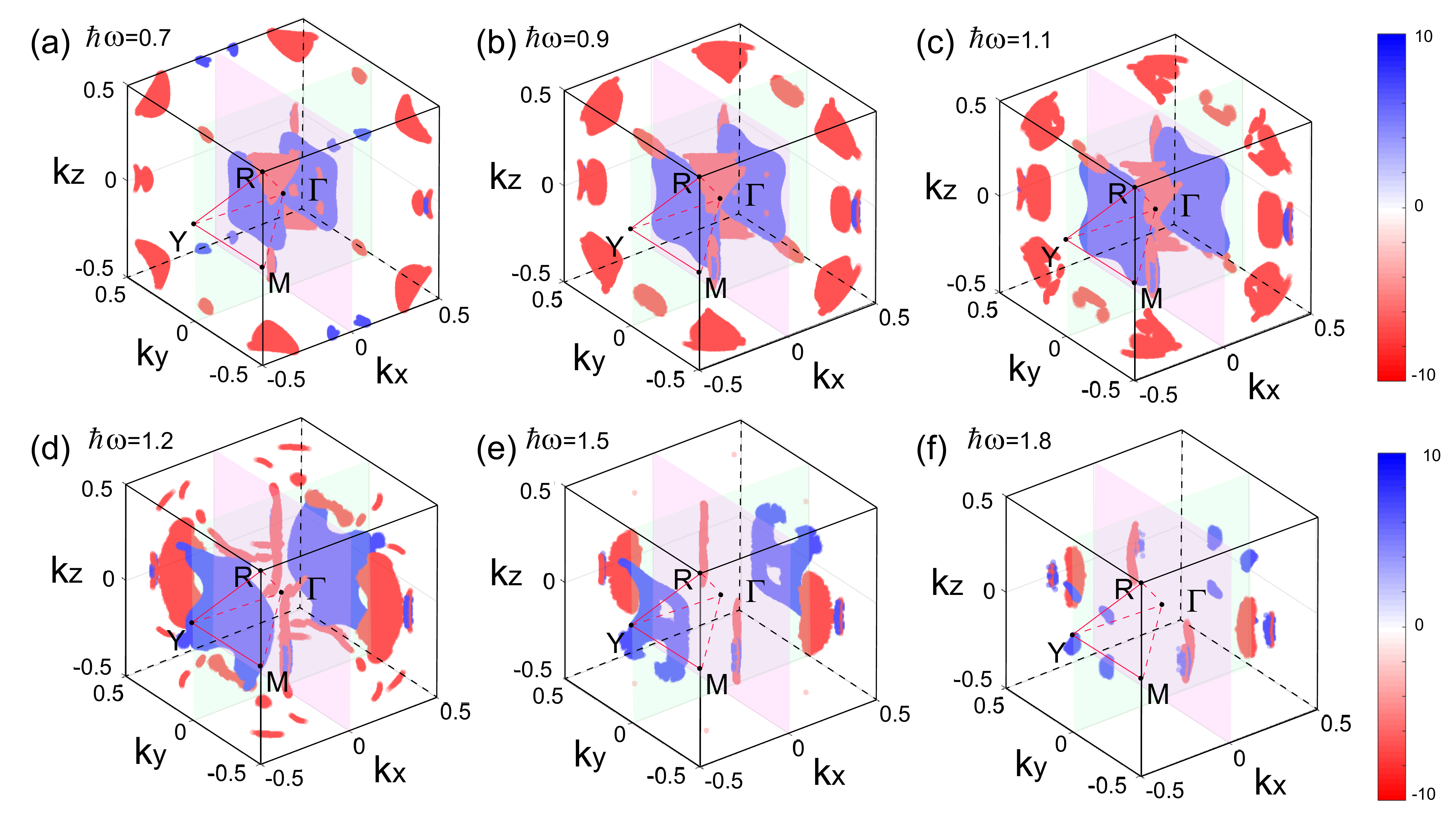}}
\caption{(color online)  Momentum distribution of CPGE tensor $\beta_{xx}$ in the Brillouin zone from all bands with corresponding Fermi level $E_{f}=0.077eV$. (a)-(f) The frequency range is from $\hbar\omega=0.7 eV$ to $\hbar\omega=1.8 eV$.
\label{fig4} }
\end{figure*}

In doping electrons, we focus on the Fermi level at fourfold degeneracy. A nearly plateau close to 4$\beta_0$ with the frequency from ~0.1 to ~0.6 eV is shown in Fig.\ref{band}(d),
where in the frequency range only optical transitions
near $\Gamma$ can contribute to CPGE trace and the
optically active fourfold fermion plays a major role. When optical
frequency $\omega$ is more than 0.6 eV, the sixfold
fermion can begin to provide CPGE trace, and cancel
the contribution from fourfold fermion, indicating
that the CPGE trace is close to zero from ~0.7 to ~2 eV.
As discussed in Section.~\ref{S2}, some contributions
to CPGE from other bands can not rule out, and the exact
quantization 4$\beta_0$ range from ~0.1 to ~0.6 eV and
exact zero rang from 0.6 to 2 eV are absent in the DFT
calculations. To analyze CPGE trace further, we
calculate the momentum distribution of CPGE
trace in the Brillouin zone in the above
different frequency ranges. Firstly,
Fig.\ref{fig3} show the momentum distribution
of CPGE trace at frequency $\hbar\omega=0.5 eV$.
Fig.\ref{fig3}(a) displays momentum distributions
from all bands with optically allowed transitions,
indicating that the CPGE completely contribute to
electric optical transitions of bands near $\Gamma$.
Momentum distributions from the band-(n-3) to band-n
are shown in Fig.\ref{fig3}(b)-(e) respectively,
where only the band-(n-2) and n are from
fourfold fermion at $\Gamma$ point and trivial
band-(n-3) and n can provide small but
finite CPGE trace shown in  Fig.\ref{fig3}(b) and
(d), suggesting that  the exact quantization
4$\beta_0$ is absent. Secondly, Fig.\ref{fig4}(a)-(f)
show momentum distributions of CPGE trace in the
Brillouin zone (BZ) from all bands with optically
allowed transitions, where the frequency range is
from $\hbar\omega=0.7 eV$ to $\hbar\omega=1.8 eV$. Range from
0.7 to 1.1 eV, the CPGE trace mainly contribute to
electric optical transitions near $\Gamma$ and $R$
points shown in Fig.\ref{fig4}(a)-(c).
Fig.\ref{fig4}(d)-(f) show the CPGE trace
of frequency from ~1.2 to ~1.8 eV, which is
provide from trivial band structures. Hence,
the CPGE trace of frequency from 0.7 to 1.8 eV
should close to zero.

\begin{figure*}
\centerline{\includegraphics[width=0.8\textwidth]{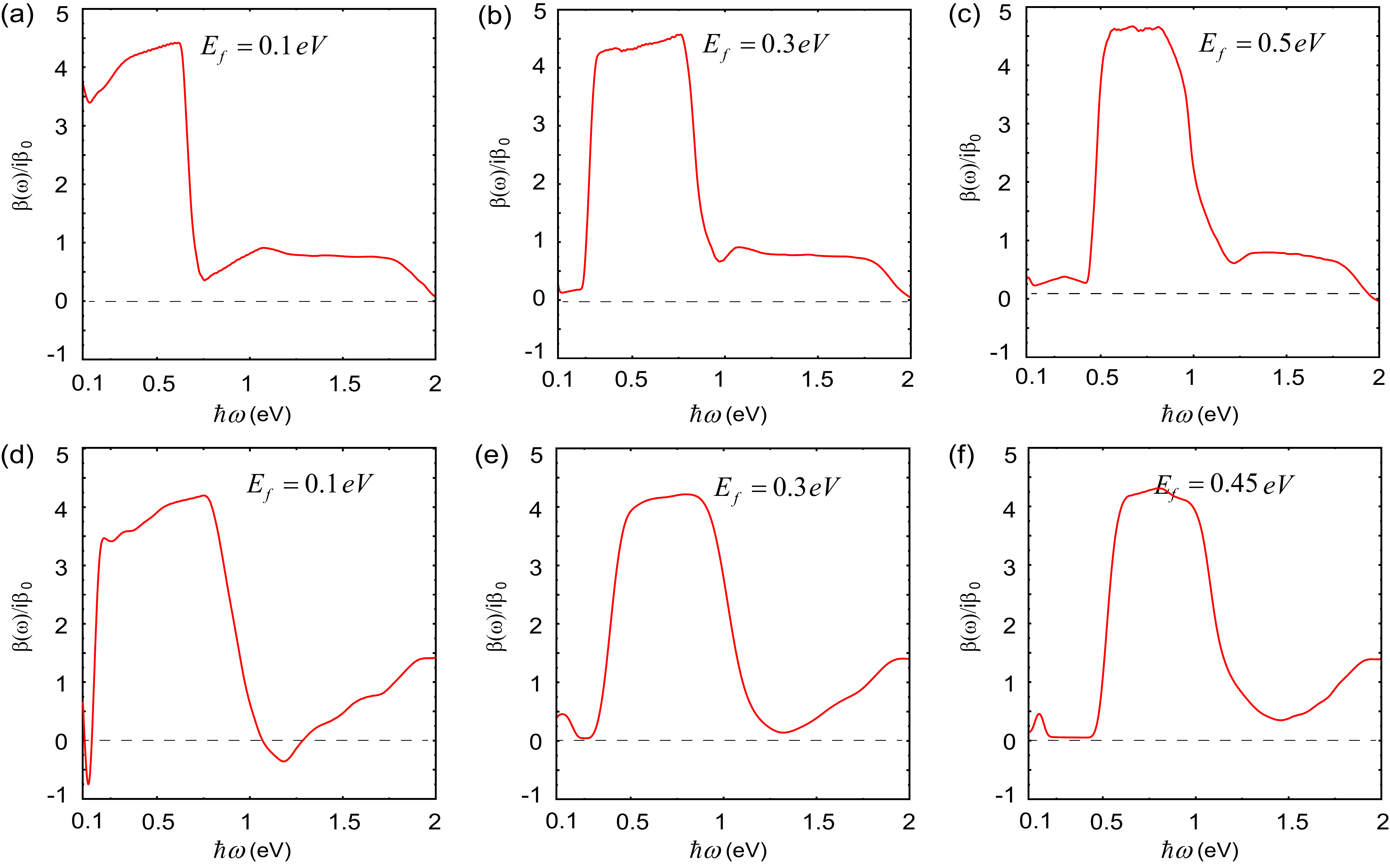}}
\caption{(color online) The trace of the CPGE tensor as  a function of frequency $\hbar\omega$ by DFT calculations with different Fermi levels (a)-(c) RhSi and (d)-(f) PtAl.
\label{PtAl} }
\end{figure*}

Except for the Fermi level at fourfold degeneracy in doping electrons, we also calculate the trace of the CPGE tensor in other Fermi levels, shown in Fig.\ref{PtAl}(a)-(c). The Fermi level from 0.1 to 0.5 $eV$, the quantized platform close to 4$\beta_0$ always exists in a certain range, which opens up more possibility for experimentally observed quantized CPGE trace in RhSi.

Since there are other related compounds(CoSi, PdGa, PtGa, PtAl, RhSn) similar to RhSi in the experiment, we also calculate the CPGE trace of these compounds. Among these compounds, PtAl is also an alternative material for obtaining quantized CPGE. In doping electrons, a quantized CPGE trace can be obtained in a certain range with the Fermi level from 0.1 to 0.45 $eV$, shown in Fig.\ref{PtAl}(d)-(f).

\section{Conclusion}\label{S4}


In summary, circular photogalvanic effect (CPGE) of the topological semimetal with chiral multifold fermions were calculated based on realistic ab-initio band structures. We find that the quantized signal is very sensitive to the trivial bands away from high symmetry points. A nearly quantized value equal to the topological charge of chiral multifold fermions can be obtained by tiny doping. When the chemical potential is around from 0.1 to 0.5 eV , A plateau close to 4 exists in the frequency range. In addition to RhSi, PtAl is also an alternative material for obtaining quantized CPGE, which appears in a certain range.

\section{ACKNOWLEDGMENTS}

This work was financially supported by the ERC Advanced Grant No. 291472 ¡®Idea Heusler¡¯, ERC Advanced
Grant No. 742068 ¡®TOPMAT¡¯. This work was performed
in part at the Center for Nanoscale Systems (CNS), a
member of the National Nanotechnology Coordinated
Infrastructure Network (NNCI), which is supported by
the National Science Foundation under NSF award no.
1541959. CNS is part of Harvard University. Some of
our calculations were carried out on the Cobra cluster of
MPCDF, Max Planck society.

\end{document}